\documentclass{article}
\usepackage{arxiv}
\usepackage{cite}
\usepackage{amsmath,amssymb,amsfonts}
\usepackage{graphicx}
\usepackage{textcomp}
\usepackage{xcolor}
\usepackage[T1]{fontenc}
\usepackage[utf8]{inputenc}
\usepackage{url}
\usepackage{hyperref}
\usepackage{microtype}
\usepackage{times}
\usepackage{booktabs}
\usepackage{listings}
\usepackage{tikz}
\usetikzlibrary{positioning,fit,backgrounds,arrows.meta,calc}
\begin{document}
\title{Evaluating Conversational Explainability in Healthcare: Comparing the Explainability Assistant and a Traditional XAI Dashboard}

\author{
Eduard Barbu\,\href{https://orcid.org/0000-0002-3664-5367}{\textnormal{\scriptsize[ORCID]}}\thanks{Corresponding author.} \\
Institute of Computer Science, University of Tartu\\
Narva mnt 18, 51009 Tartu, Estonia\\
\texttt{eduard.barbu@ut.ee}
\And
Rodion Krjut\v{s}kov \\
Nupp Software\\
\~{O}ism\"{a}e tee 25-59, 13514 Tallinn, Estonia\\
\texttt{rodion@nupp.dev}
\And
Anneli Uusk\"{u}la \\
Institute of Family Medicine and Public Health\\
Ravila 19, 50411 Tartu, Estonia\\
\texttt{anneli.uuskula@ut.ee}
}

\date{28 October 2026}
\maketitle

\begin{center}\small Preprint prepared for EXPLIMED 2026.\end{center}

\begin{abstract}
Conversational interfaces have been proposed as a more accessible
alternative to dashboard-based explainable AI (XAI) tools for users
without machine learning expertise. In healthcare, where predictive
models are increasingly deployed for clinical decision support and
end-users span a wide range of technical backgrounds, the practical
benefits of conversational explanation remain insufficiently
characterised. Existing healthcare conversational XAI systems have
typically been evaluated in isolation rather than against the
dashboard interfaces they aim to displace, leaving open the question
of whether the conversational modality itself contributes to
interpretability beyond the underlying explanation methods.

This paper reports a within-subjects user study comparing the
Explainability Assistant, a conversational XAI system based on LLM
function-calling, with a traditional dashboard interface for the
explanation of a heart disease classification model trained on the
Cleveland Heart Disease dataset. Both interfaces provide access to
the same Decision Tree classifier and the same post-hoc explanation
methods (SHAP feature attribution, DiCE counterfactuals, and what-if
analysis), so that observed differences can be attributed to the
interaction modality rather than to differences in explanation
content. Five participants with healthcare backgrounds, ranging from
medical students to practising clinicians and public health
researchers, completed parallel task blocks on each interface; three
also took part in follow-up interviews. Quantitative measures (task
accuracy and Likert ratings) and qualitative analyses (free-text
questionnaire responses and interview transcripts) converge on a
nuanced finding: rather than clearly preferring one modality over
the other, participants identified complementary strengths and
weaknesses on both sides, and recommended an integrated tool
combining conversational and dashboard elements in a single clinical
workflow.
\end{abstract}

\keywords{Explainable AI \and
  Healthcare \and
  Conversational interfaces \and
  Clinical decision support \and
  Human-centered XAI \and
  User study}

\section{Introduction}
\label{sec:introduction}

Machine learning models are increasingly deployed in healthcare for
clinical decision support, from triage and risk prediction to
diagnostic assistance and treatment planning. Predictive accuracy
alone, however, is rarely sufficient for adoption in a clinical
workflow: practitioners must understand why a model has produced a
given prediction in order to assess its plausibility against
clinical evidence, communicate it to colleagues and patients, and
intervene when the model is wrong. The post-hoc explanation methods
that have become standard in this setting (SHAP attribution
\cite{lundberg2017}, counterfactual generation \cite{mothilal2020},
and what-if exploration) address part of this requirement by
attaching interpretable structure to individual predictions, but
they leave a separate question open: in what form should these
explanations be delivered to clinical end-users with widely varying
levels of machine learning expertise?

The dominant interface paradigm has been the explanation dashboard,
a graphical environment that consolidates the available explanation
methods into a single point-and-click application. Dashboards are
information-dense and well-suited to systematic inspection, but
they presuppose that the user can map a clinical question onto the
appropriate dashboard view, which is a non-trivial cost for
clinical end-users without machine learning training. Conversational
interfaces driven by large language models (LLMs) have recently
been proposed as a more accessible alternative
\cite{slack2023,samimi2025}, on the premise that allowing users to
express explanation needs in natural language removes the mapping
step and lowers the cognitive cost of using the underlying methods.
Evaluations of conversational XAI in healthcare, however, have
typically reported on the conversational system in isolation rather
than against the dashboard interfaces it is meant to displace,
leaving open the question of whether the conversational modality
itself contributes to interpretability beyond the contribution of
the explanation methods themselves.

This paper reports a within-subjects user study designed to
isolate that contribution. The Explainability Assistant \cite{krjutskov2026explainability}, a
conversational XAI system based on LLM function-calling, is
compared against a traditional dashboard interface for the
explanation of a Decision Tree classifier trained on the Cleveland
Heart Disease dataset. Both interfaces expose the same classifier
and the same set of post-hoc explanation methods (SHAP, DiCE, and
what-if analysis), so that observed differences in task accuracy,
subjective ratings, and qualitative response patterns can be
attributed to the modality of interaction rather than to
differences in explanation content. Five participants with mixed
healthcare backgrounds, spanning practising clinicians, medical
and nursing students, and public health researchers, completed
parallel task blocks on both interfaces, and three of these
participants took part in follow-up interviews. The analysis
combines descriptive quantitative comparison with two layers of
qualitative evidence: the free-text responses participants gave
alongside the task questions and in the final open-ended item, and
the recorded interview discussions.

The principal finding of the study cuts against the prevailing
expectation from earlier conversational XAI work. Rather than
selecting one modality over the other, participants identified
complementary strengths and weaknesses on both sides: the
conversational assistant offered easier entry and clearer scoping
of responses to individual patients, while the dashboard offered
simultaneous visibility of global and local information and the
visible separation of the underlying model from the conversational
layer. All three interview participants explicitly recommended an
integrated tool combining the two modalities in a single clinical
workflow.

The remainder of the paper is organised as follows.
Section~\ref{sec:related-work} reviews related work on clinical
XAI, conversational interfaces, and trust calibration.
Section~\ref{sec:system} describes the Explainability Assistant
architecture, and Section~\ref{sec:task} the prediction task and
the explanation methods shared by the two interfaces.
Section~\ref{sec:llm-eval} reports a benchmark evaluation of six candidate LLMs against the tool-calling requirements imposed by the architecture. Section~\ref{sec:user-study} describes the user
study, Section~\ref{sec:results} reports the quantitative and
qualitative results, and Sections~\ref{sec:discussion}
and~\ref{sec:conclusion} discuss the findings and conclude.

\section{Related Work}
\label{sec:related-work}

The work presented here connects three strands of prior research:
explanation accessibility for non-technical users in clinical
decision support, conversational interfaces for model
interpretability, and the empirical literature on trust calibration
in human-AI collaboration.

Post-hoc explainability methods such as SHAP \cite{lundberg2017},
LIME \cite{ribeiro2016}, and counterfactual approaches
\cite{mothilal2020} are standard tools for attributing predictions
in clinical machine learning models, but their direct presentation
to clinical end-users is a known source of friction. A controlled
study with 63 surgeons and physicians on pre-operative blood product
prescription found that clinician-friendly explanation formats
produced higher acceptance, trust, and usability scores than raw
SHAP visualisations, with LLM-generated narrative explanations
emerging as a candidate solution to the clarity gap
\cite{Hur2025SHAP}. Critical perspectives on medical XAI argue
that current methods are calibrated to the needs of model developers
rather than clinical end-users and that individual feature
attributions may not constitute clinically meaningful explanations
\cite{ghassemi2021,tonekaboni2019}. Interactive dashboards such as
the What-If Tool \cite{wexler2020}, the Language Interpretability
Tool \cite{tenney2020}, and the open-source Explainer Dashboard \cite{oege_dijk_2023_7633294}
consolidate multiple explanation methods into a single graphical
interface, but presuppose that the user can map a clinical question
onto the appropriate dashboard view, a non-trivial cost for users
without machine learning training that often shifts the interaction
from active inquiry to passive inspection.

Natural-language dialogue lowers this accessibility barrier by
allowing users to express explanation needs directly. TalkToModel
\cite{slack2023} demonstrated the paradigm using fine-tuned T5
models with custom semantic grammars, reporting maximum utterance parsing accuracy of
76.8\%. More recent designs leverage the function-calling
capabilities of frontier models to obtain structured tool
invocations without task-specific fine-tuning, while approaches such
as LLMCheckup \cite{wang2024} rely on prompt chaining over
general-purpose LLMs. A systematic review of dialogue-based XAI
\cite{mindlin2025} notes the relative scarcity of comparative
evaluations against non-conversational baselines. The most direct
healthcare-domain comparator is the visual-conversational interface
for diabetes risk prediction of Samimi et al.\ \cite{samimi2025},
which combines a fine-tuned T5 parser with a general-purpose LLM
and grounds explanations in scientific evidence. A mixed-methods
study with 30 healthcare professionals reported that conversational
interaction supported clearer understanding of model assessments
and that evidence grounding contributed to trust calibration. The
system retains a fine-tuned analytical component, and the
evaluation does not include a head-to-head comparison against a
dashboard interface providing the same underlying explanation
methods.

Explanation interfaces in clinical settings should also support
appropriate reliance: users should follow the model's recommendation
when it is correct and override it when it is not
\cite{schemmer2023}. Empirical evidence on whether explanations
achieve this is mixed. A study with 223 dermatologists on lesion
classification reported that self-reported trust accounted for up
to 24\% of the variance in reliance behaviour, with both
over-reliance and under-reliance observed at the extremes of the
trust distribution \cite{kuper2025}. Reviews of human-AI
collaboration in high-stakes domains have documented that
explanation provision can promote over-reliance through
false-confirmation effects \cite{informatics12040135}, and the risk that
conversational interfaces may amplify over-reliance through the
persuasive fluency of natural-language responses has been raised
in recent dialogue-based XAI work \cite{he2025}. These findings
motivate evaluation designs that measure objective task accuracy
alongside subjective trust ratings.

Against this background, the present paper compares the
Explainability Assistant, a conversational XAI system that uses LLM
function-calling for intent parsing and analytical tool invocation,
against a dashboard baseline using the same classifier and the same
post-hoc explanation methods, so that observed differences can be
attributed to the interaction modality itself rather than to
differences in explanation content. The participant pool spans
practising clinicians, medical and nursing students, and public
health researchers, providing a broader test of accessibility than
evaluations restricted to a single professional category. Both
quantitative measures and qualitative data are collected, the
latter including free-text questionnaire responses and a subset of
follow-up interviews.

\section{System Overview}
\label{sec:system}

The Explainability Assistant \cite{krjutskov2026explainability} is a conversational interface to a
configured set of post-hoc explanation methods over a fixed
predictive model. Its application logic is generic across deployment
domains: a deployment is specified through configuration files that
fix the dataset, the trained model artefact, the JSON function
specifications exposing the explanation methods, and the system
prompt under which the LLM operates. The heart disease deployment
evaluated in this paper is obtained through configuration alone, as
detailed in Section~\ref{sec:task} and
Section~\ref{sec:heart-adaptation}. The remainder of this section
describes the architecture, the function-calling pipeline, and the
response contract that governs information flow between the back-end
and the LLM provider.

\subsection{Architecture}

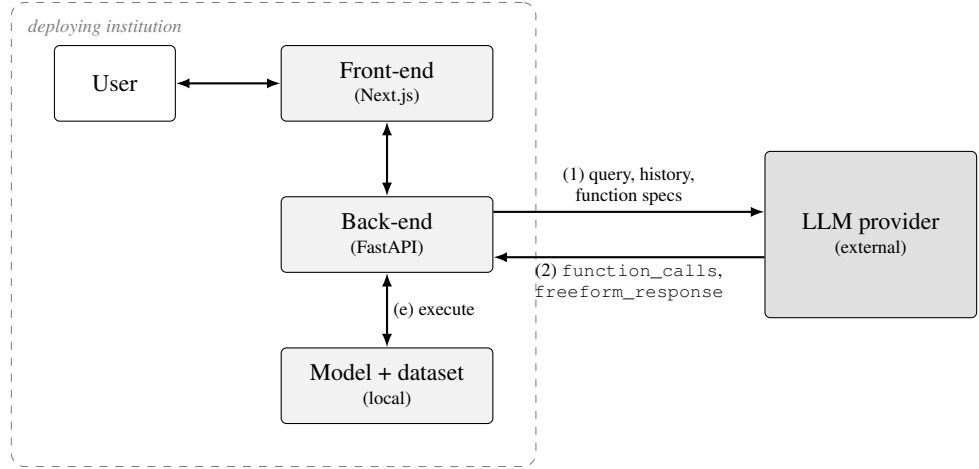
\begin{figure}[t]
  \centering
  \begin{tikzpicture}[
    >={Latex[length=1.8mm]},
    box/.style={
      draw, rounded corners=2pt, align=center,
      font=\small, inner sep=4pt, minimum height=10mm
    },
    component/.style={box, fill=black!5, minimum width=28mm},
    user/.style={box, fill=white, minimum width=16mm},
    external/.style={box, fill=black!12, minimum width=28mm, minimum height=22mm},
    lbl/.style={font=\scriptsize, align=center, inner sep=2pt}
  ]
    \node[user]      (user)  at (0, 0)      {User};
    \node[component] (front) at (3.6, 0)    {Front-end\\[-1pt]\scriptsize (Next.js)};
    \node[component] (back)  at (3.6, -2.0) {Back-end\\[-1pt]\scriptsize (FastAPI)};
    \node[component] (model) at (3.6, -4.0) {Model + dataset\\[-1pt]\scriptsize (local)};
    \node[external]  (llm)   at (10.0, -2.0) {LLM provider\\[-1pt]\scriptsize (external)};

    \begin{scope}[on background layer]
      \node[draw, dashed, rounded corners=5pt, gray,
            fit=(user)(front)(back)(model), inner sep=16pt] (tb) {};
    \end{scope}
    \node[font=\scriptsize\itshape, gray, anchor=north west]
      at ([xshift=3pt, yshift=-3pt]tb.north west) {deploying institution};

    \draw[<->, thick] (user)  -- (front);
    \draw[<->, thick] (front) -- (back);
    \draw[<->, thick] (back)  -- node[lbl, right] {(e) execute} (model);

    \draw[->, thick] ([yshift=3mm]back.east) --
      node[lbl, above] {(1) query, history,\\function specs}
      ([yshift=3mm]llm.west);
    \draw[->, thick] ([yshift=-3mm]llm.west) --
      node[lbl, below] {(2) \texttt{function\_calls},\\\texttt{freeform\_response}}
      ([yshift=-3mm]back.east);
  \end{tikzpicture}
  \caption{System architecture and response contract. Tool outputs
    are rendered through static templates and are not returned to
    the LLM at any point in the interaction.}
  \label{fig:architecture}
\end{figure}

\begin{figure}
    \centering
    \includegraphics[width=0.75\linewidth]{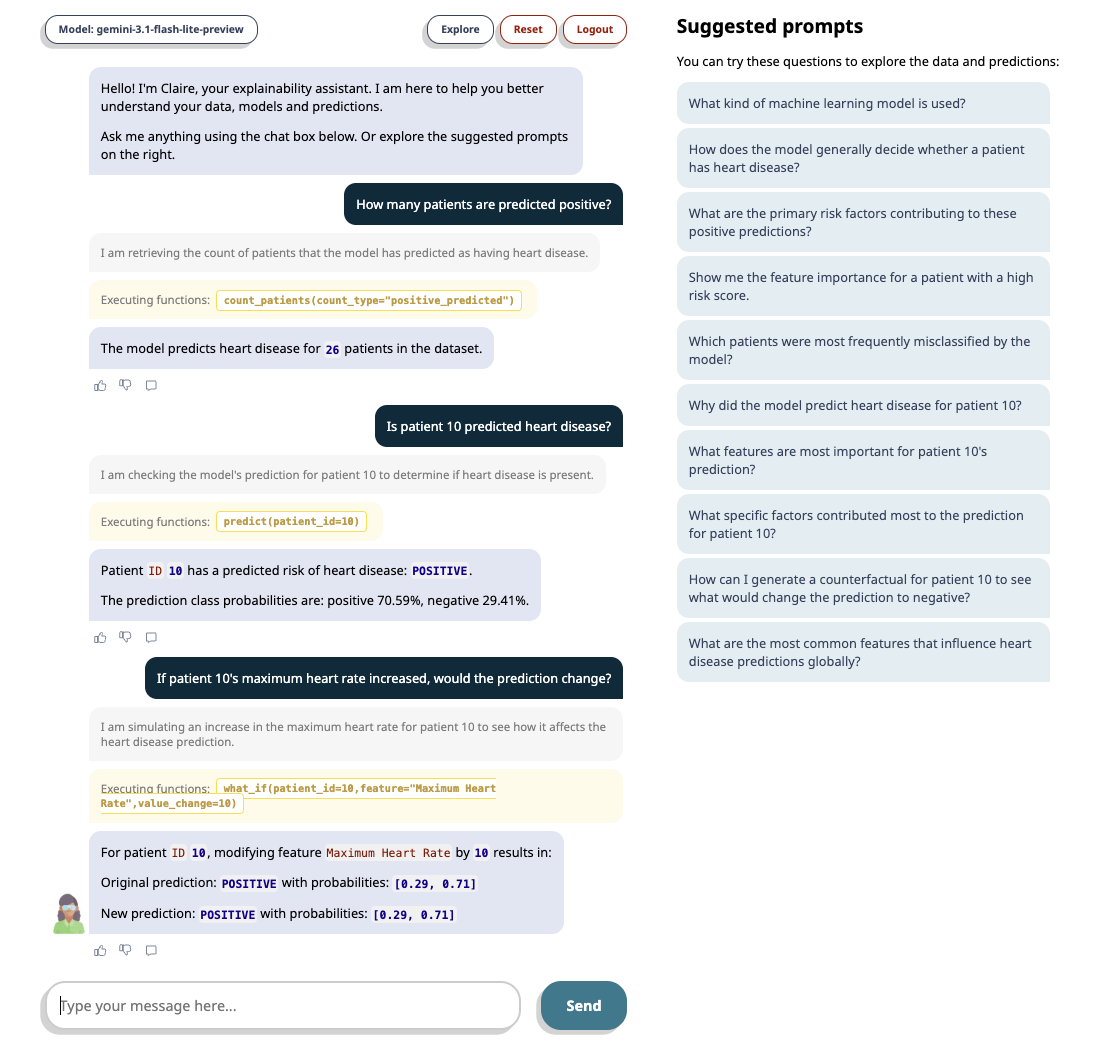}
    \caption{Explainability Assistant user interface. Each response contains up to three parts: agent reasoning trace, list of tools used and a rendering of a static template. Suggested prompts are generated by a separate agent based on the state of the conversation.}
    \label{fig:ea-ui}
\end{figure}

The system comprises a Next.js front-end and a FastAPI back-end,
with a separation between presentation and computational layers
(Figure~\ref{fig:architecture}). The front-end (Figure~\ref{fig:ea-ui}) manages the
conversation history visible to the user, allows selection among
supported LLM models, and renders both the assistant's free-form
reasoning and the structured outputs of executed functions. The back-end
orchestrates model inference, executes the configured explanation,
prediction and data inspection methods, and communicates with external LLM providers
through respective APIs. No long-lived server-side session state is required: the
conversation history is supplied with each request, and the
back-end is otherwise stateless. The current deployment supports
several LLM providers behind a uniform internal interface, including
instances from the Gemini, GPT, and open-weight families through OpenRouter.
Adding a new provider amounts to implementing the internal interface and
registering it in the back-end configuration; no other changes to
the application logic are required, which allows the same heart
deployment to be benchmarked across models and providers as reported in
Section~\ref{sec:llm-eval}.

\subsection{Function-Calling Pipeline}

For every user message, the back-end constructs a request to the
LLM containing a subset of conversation history, a system prompt describing
the agent's role, and a JSON specification of the available
functions. Each specification entry lists the function name, a
natural-language description of when it should be used, and the
schema of its arguments. Modern LLMs with function-calling support
map the user's request onto one or more of these functions and
populate the arguments from the conversation context, without task-specific
fine-tuning. The functions exposed by the back-end implement the
post-hoc explanation methods described in Section~\ref{sec:task}
(SHAP feature attribution, DiCE counterfactual generation, what-if
analysis), together with auxiliary operations supporting typical
conversational queries: retrieval of an individual patient record,
computation of model performance metrics, identification of
misclassified cases, retrieval of dataset and model summaries, and
counts of patients in clinically defined subgroups. The full
catalogue used for the heart deployment contains twenty functions
and is published with the source code.

This design contrasts with prompt-chaining approaches such as
LLMCheckup \cite{wang2024}, in which the LLM mediates the
orchestration of multiple intermediate prompts, and with
grammar-based parsers such as TalkToModel \cite{slack2023}, in
which a fine-tuned analytical model emits a custom intermediate
representation of the query. Here, the function specifications are the only
contract between the LLM and the back-end, and the LLM is treated
as an off-the-shelf component restricted to intent parsing and
argument selection.

\subsection{Zero-Leakage Response Contract}
\label{sec:zero-leakage}

A clinical deployment must operate under tighter constraints on
the flow of patient information than a deployment in a less
sensitive domain. Patient-level records and the outputs of
explanation methods over those records may contain or reconstruct
protected health information, and once such information enters an
LLM context window operated by an external provider, the controls
available to the deploying institution are limited to whatever the
provider's data handling policy and contractual terms allow. The
design adopted here removes the question by eliminating the
channel: tool outputs are not passed back to the LLM at any point
in the interaction, and any information the user sees about a
tool's result is rendered through pre-authored templates over the
structured output of the function rather than through a generative
summary.

The response the LLM is required to produce on each turn is
constrained by the schema illustrated in Figure~\ref{fig:schema}
to two fields. The \emph{function\_calls} field lists the function
invocations the assistant proposes to execute, expressed as
ordinary call strings rather than as native provider tool-call
objects so that the schema is uniform across providers. The
\emph{freeform\_response} field contains the agent's
natural-language description of what it is about to do, prepared
without sight of any tool output and shown to the user alongside,
but separately from, those outputs once they are computed. The
back-end parses the function calls, executes each function locally
against the trained model artefact and the dataset, and dispatches the
structured results to the front-end for rendering. The LLM is not
consulted again within the same turn and never receives the tool
outputs.

\begin{figure}[t]
  \centering
  \begin{tikzpicture}[
    panel/.style={
      draw, rounded corners=2pt, fill=black!5,
      inner sep=6pt, font=\small, align=left,
      text width=82mm
    },
    note/.style={font=\scriptsize\itshape, align=left, text width=82mm}
  ]
    \node[panel] (schema) {%
      \textbf{Response schema (LLM-emitted):}\\[3pt]
      \quad \texttt{function\_calls}: list of call strings\\[2pt]
      \quad \texttt{freeform\_response}: single string
    };

    \node[panel, below=3mm of schema] (example) {%
      \textbf{Example turn:}\\[3pt]
      \emph{User:} ``What did the model predict for patient 42, and which features contributed most?''\\[3pt]
      \emph{LLM response:}\\[1pt]
      {\footnotesize\ttfamily
      function\_calls: [\\
      \hspace*{1em}predict(patient\_id=42),\\
      \hspace*{1em}feature\_importance\_patient(patient\_id=42)\\
      ]\\[2pt]
      freeform\_response:\\
      \hspace*{1em}\rmfamily ``I will retrieve the prediction for patient 42 and the SHAP attributions explaining it.''}
    };

  \end{tikzpicture}
  \caption{Required response schema. The LLM emits the proposed
    function calls and a free-form description of its intended
    actions; tool outputs are produced by the back-end and do not
    enter the response.}
  \label{fig:schema}
\end{figure}

This contract has three properties relevant to a clinical
deployment. First, data minimisation at the LLM boundary: the only
material transmitted to the provider is the user's natural-language
request, the limited conversation history including user requests
and corresponding function call strings, along with the function specifications;
patient records, computed feature attributions, and counterfactual examples remain
on the back-end. Second, non-fabrication of tool output: the
assistant cannot paraphrase or alter the result of an explanation
method because it does not see it, and the user sees the
deterministic output of the method as rendered by the front-end.
Third, interpretability of intent: the user sees the listed
function calls and the free-form description together, and a
misinterpretation of the request is recoverable by reformulating
the question rather than by inspecting opaque output. The term
\emph{Zero-Leakage} is used in the remainder of the paper to refer
to the configuration in which this contract is enforced; it is
taken as the architectural baseline against which the evaluations
in Sections~\ref{sec:llm-eval} and \ref{sec:user-study} are
conducted.

\subsection{Heart Disease Adaptation}
\label{sec:heart-adaptation}

The heart disease deployment of Explainability Assistant is obtained from the system
configuration through changes to a small number of files, with no modification
to the core application logic. The dataset and trained model
artefacts described in Section~\ref{sec:task} are loaded by the
back-end; the JSON function specifications expose the explanation
methods over the trained model artefact and use the clinical feature names of
the Cleveland dataset (\emph{trestbps}, \emph{thalach},
\emph{oldpeak}, and so on); and the system prompt refers to clinical context.
The set of explanation methods, the response schema, the front-end rendering, and the
Zero-Leakage contract are largely unchanged from previously explored use cases.
Where a feature name is opaque to a non-specialist user, for instance \emph{ca} for the
number of major vessels coloured by fluoroscopy, the function
specification surfaces the longer form so that the LLM can use
it in its free-form response and the front-end can use it in
template labels. A separate agent operating under the same response contract 
was introduced to the system to generate suggested 
prompt follow-ups based on the state of the conversation.

\section{Healthcare Prediction Task}
\label{sec:task}

The prediction task selected for this study is the diagnosis of
coronary artery disease from a small set of routinely collected
clinical attributes. The task is well-known in the medical machine
learning literature, the underlying dataset is publicly available,
and the clinical features have established interpretations
familiar to participants with healthcare backgrounds, making it a
suitable testbed for an evaluation focused on the interpretability
of model outputs rather than on the modelling problem itself.

\subsection{Dataset}

The Cleveland Heart Disease dataset \cite{detrano1989} is used in
its standard form as distributed by the UCI Machine Learning
Repository\footnote{\url{https://archive.ics.uci.edu/dataset/45/heart+disease}}.
It contains records for 303 patients evaluated at the Cleveland
Clinic Foundation, with 14 attributes per record: thirteen
predictors and one target. The predictor variables include patient
demographics, symptoms reported during clinical interview, and the
results of standard non-invasive examinations such as resting
electrocardiography and exercise stress testing
(Table~\ref{tab:features}).

\begin{table}
  \caption{Predictor variables in the Cleveland Heart Disease dataset.}
  \label{tab:features}
  \begin{tabular}{lll}
    \toprule
    Variable & Type & Description \\
    \midrule
    age & numeric & Age in years \\
    sex & categorical & Patient sex \\
    cp & categorical & Chest pain type (four levels) \\
    trestbps & numeric & Resting blood pressure (mm Hg) \\
    chol & numeric & Serum cholesterol (mg/dl) \\
    fbs & binary & Fasting blood sugar $>$ 120 mg/dl \\
    restecg & categorical & Resting electrocardiographic results \\
    thalach & numeric & Maximum heart rate achieved \\
    exang & binary & Exercise-induced angina \\
    oldpeak & numeric & ST depression induced by exercise \\
    slope & categorical & Slope of the peak exercise ST segment \\
    ca & numeric & Number of major vessels coloured by fluoroscopy \\
    thal & categorical & Thalassemia status \\
    \bottomrule
  \end{tabular}
\end{table}

The original target variable encodes coronary artery disease
severity on a five-point ordinal scale (0-4) corresponding to the
proportion of major vessels exhibiting more than 50\% diameter
narrowing on angiography. Following established practice, the
target is dichotomised: 0 indicates the absence of disease, and
any value greater than 0 indicates its presence. This formulation
reflects the clinically actionable distinction between patients
who do and do not require further investigation and produces a
roughly balanced class distribution (approximately 54\% negative,
46\% positive). Missing values for \emph{ca} and \emph{thal},
encoded as the literal character \texttt{?}, are parsed accordingly,
and complete-case filtering leaves 297 patients in the working
dataset. Categorical features are encoded as integers preserving
the original UCI codings, and numeric features are retained on
their natural scales so that values displayed to participants
remain in clinically recognisable units.

\subsection{Classifier and Explanation Methods}

A Decision Tree classifier is used as the predictive model. The
choice is motivated by fit between model and study purpose rather
than maximisation of predictive performance: decision trees are
intrinsically interpretable, providing a clear ground truth
against which the plausibility of the post-hoc explanations can
be assessed informally, while the explanation methods themselves
(SHAP, DiCE, and what-if analysis) are model-agnostic and apply
unchanged as they would to a more opaque classifier.

The classifier is implemented using \texttt{scikit-learn}
\cite{pedregosa2011} with maximum depth 4, minimum five samples
per leaf, and a fixed random seed; all other hyperparameters
retain library defaults. Performance is evaluated on a stratified
80/20 holdout split (237 training, 60 test). The data preparation
script, the trained model artefact, and the evaluation outputs
are available in a public repository\footnote{Repository URL
withheld for double-blind review; will be added at camera-ready.}.
On the held-out test partition the classifier achieves an accuracy
of 80\% and a weighted F1 score of 0.80
(Table~\ref{tab:dt-performance}). Class-conditional performance is
moderately asymmetric: precision and recall for the negative class
are 0.79 and 0.84, whereas for the positive class they are 0.81
and 0.75. The lower recall for the positive class indicates that
the model is somewhat more likely to miss disease cases than to
flag healthy patients incorrectly, an asymmetry visible in the
misclassified cases the model can be asked about through either
interface. The performance level is
sufficient to support the explanation study, in which predictive
accuracy is not the variable of interest.

\begin{table}
  \caption{Decision Tree classifier performance on the held-out
    test set (60 patients).}
  \label{tab:dt-performance}
  \begin{tabular}{lccc}
    \toprule
    Class & Precision & Recall & F1 \\
    \midrule
    No disease (0) & 0.79 & 0.84 & 0.82 \\
    Disease (1)    & 0.81 & 0.75 & 0.78 \\
    \midrule
    Macro average    & 0.80 & 0.80 & 0.80 \\
    Weighted average & 0.80 & 0.80 & 0.80 \\
    \bottomrule
  \end{tabular}
\end{table}

Both the Explainability Assistant and the dashboard baseline
expose the same set of post-hoc explanation methods, configured
identically: SHAP \cite{lundberg2017} for local and global feature
attribution; DiCE \cite{mothilal2020} for counterfactual
generation, configured to respect training-data feature ranges so
that suggested modifications correspond to clinically plausible
patient profiles; and user-directed what-if analysis in which the
user specifies feature modifications and observes the resulting
change in prediction. This identity of explanation methods is
essential to the study design: any difference in observed outcomes
can be attributed to the modality of interaction rather than to
differences in explanation content.

\section{LLM Tool-Calling Evaluation}
\label{sec:llm-eval}

The Zero-Leakage contract imposes a strict requirement on the LLM:
the model must produce a JSON object containing the list of
function calls and a free-form description of the planned actions,
with both fields valid for the back-end to proceed. The model
never receives tool outputs and cannot recover from an error in
argument selection by reading the result of a call. The
architecture is therefore feasible only to the extent that
available LLMs can produce the required object reliably under the
request distribution generated by clinical end-users. This section
reports an evaluation of six LLMs on a held-out tool-calling
benchmark over the heart deployment, together with a complementary
evaluation of free-form response quality on cases where the
correct behaviour is to refrain from calling a tool.

\subsection{Evaluation Setup}

A benchmark of 343 conversational turns was constructed for the
heart function catalogue described in Section~\ref{sec:system}.
The benchmark exercises every supported tool together with ten
scenario types covering direct single-turn requests, multi-tool
requests, references to entities established earlier in the
conversation, paraphrased and aliased queries, requests with
missing arguments, conversational turns that should not produce a
tool call, requests that conflict with the conversation context,
irrelevant context, requests the catalogue does not support, and
switches or corrections of the entity under discussion. Scenario
sizes range from 14 cases (irrelevant context) to 98 cases (direct
single-turn requests); the distribution is reported in
Table~\ref{tab:scenario-breakdown}.

The dataset was assembled in three stages. A small seed of
manually authored cases established trusted labels for each
scenario type. A teacher model Claude Opus 4.7 \cite{opus47}
(a frontier LLM that excels at knowledge work, not included in the
evaluation) was used to scale the seed to a larger candidate pool
by analogy with the seed examples. Every teacher-generated case was 
then pre-screened by GPT-5.5 \cite{gpt55} (not included in the
evaluation) and subsequently reviewed and corrected by hand before 
being admitted to the final evaluation set, on the principle that admitting raw teacher
output would risk evaluating whether other models imitate the
teacher's tool selection rather than whether they follow the heart
catalogue correctly.

Six models were evaluated in full: Gemma~4 Mixture of Experts,
Gemma~4 (dense) \cite{gemma4}, Gemini~3.1 Flash Lite \cite{gemini31},
DeepSeek V4 Flash \cite{deepseek4}, GPT-5.4-mini \cite{gpt54mini},
and Kimi K2.5 \cite{kimi25}. A seventh candidate, Qwen~3.6 Flash \cite{qwen36},
was smoke-tested on ten cases and excluded because of consistent response-contract
failure. All six evaluated models were
operated through the same back-end path used by the live
application; native provider tool-calling extensions were not used,
and the response object was returned as ordinary structured JSON.

Prediction scoring canonicalises arguments before exact-match
comparison so that semantically equivalent calls are not penalised
for surface variation; feature aliases, what-if values expressed
in alternative forms, and metric aliases are compared by their
canonical back-end meaning. The reported metrics are: response
schema validity (RSV), function-call string validity (FCV), intent
accuracy (IA), argument accuracy (AA), joint goal accuracy (JGA),
no-call accuracy (NCA), and the rates of overcall, undercall, and
hallucinated tools (calls to functions absent from the catalogue).

\subsection{Tool-Calling Accuracy}

Aggregate results across the six models are reported in
Table~\ref{tab:tool-calling-overall}. All six satisfy the response
schema on essentially every case. Function-call string validity is
similarly high. The substantive differentiation lies in intent and
argument accuracy, joint goal accuracy, and no-call discipline.

\begin{table}
  \caption{Aggregate tool-calling metrics on the reviewed dataset
    (343 cases). Best values in bold.}
  \label{tab:tool-calling-overall}
  \small
  \begin{tabular}{lccccccccc}
    \toprule
    Model & RSV & FCV & IA & AA & JGA & NCA & Over & Under & Hall \\
    \midrule
    Gemma~4 MoE      & 1.000 & 0.997 & 0.898 & 0.908 & \textbf{0.898} & \textbf{0.918} & \textbf{0.023} & 0.058 & 0.003 \\
    Gemma~4 dense    & 1.000 & 1.000 & 0.898 & 0.961 & 0.895 & 0.753 & 0.070 & 0.003 & 0.000 \\
    Gemini~3.1~flash       & 1.000 & 1.000 & 0.875 & \textbf{0.965} & 0.869 & 0.670 & 0.093 & 0.003 & 0.000 \\
    DeepSeek V4      & 1.000 & 0.983 & 0.837 & 0.898 & 0.828 & 0.711 & 0.082 & 0.017 & 0.017 \\
    GPT~5.4~mini     & 1.000 & 0.980 & 0.802 & 0.855 & 0.790 & 0.680 & 0.090 & 0.038 & 0.020 \\
    Kimi K2.5        & 0.985 & 0.971 & 0.729 & 0.731 & 0.720 & 0.691 & 0.079 & 0.096 & 0.015 \\
    \bottomrule
  \end{tabular}
\end{table}

The two Gemma variants and Gemini lead by joint goal accuracy:
Gemma~MoE (0.898), Gemma~dense (0.895), and Gemini (0.869), with
DeepSeek (0.828), GPT (0.790), and Kimi (0.720) trailing. The
leaders' advantage is built differently. Gemma~dense and Gemini
achieve their scores through high argument accuracy (0.961 and
0.965) but tolerate a higher rate of overcalls in ambiguous
contexts; Gemma~MoE has a markedly lower overcall rate (0.023
against 0.070 and 0.093) and the strongest no-call accuracy
(0.918), at the cost of weaker carryover and paraphrase handling.

The scenario-level breakdown in Table~\ref{tab:scenario-breakdown}
clarifies this trade-off. Gemma~MoE is the strongest model on
scenarios that reward conservative calling: conflicting context
(0.867 against 0.600 for the next model), unsupported intent
(0.920), and multi-tool request (0.960). Gemma~dense and Gemini
lead on scenarios that reward stateful precision: parameter
carryover (0.947 for both, against 0.711 for MoE), paraphrase or
alias (0.970 for both), and entity switch or correction. The two
patterns describe two coherent strategies: a conservative model
that requests clarification under ambiguity but loses some
supported tool executions, and a forward model that resolves
ambiguous contexts by selecting the most plausible referent and
is rewarded on supported cases but penalised on ambiguous ones.

\begin{table}
  \caption{Joint goal accuracy by scenario. Cases column gives the
    number of benchmark cases per scenario.}
  \label{tab:scenario-breakdown}
  \small
  \begin{tabular}{lccccccc}
    \toprule
    Scenario & Cases & MoE & Dense & Gemini & DeepSeek & GPT & Kimi \\
    \midrule
    conflicting context        & 15 & \textbf{0.867} & 0.200 & 0.133 & 0.533 & 0.600 & 0.400 \\
    direct single turn         & 98 & 0.939 & \textbf{0.949} & 0.939 & 0.867 & 0.888 & 0.704 \\
    entity switch / correction & 38 & 0.895 & \textbf{0.974} & 0.921 & 0.947 & 0.895 & 0.737 \\
    irrelevant context         & 14 & \textbf{1.000} & 0.929 & \textbf{1.000} & 0.929 & 0.929 & 0.929 \\
    missing required argument  & 34 & \textbf{0.912} & \textbf{0.912} & 0.824 & 0.618 & 0.500 & 0.647 \\
    multi-tool request         & 25 & \textbf{0.960} & 0.920 & \textbf{0.960} & 0.920 & 0.840 & 0.840 \\
    no tool needed             & 23 & 0.957 & 0.913 & 0.826 & 0.957 & 0.913 & \textbf{1.000} \\
    parameter carryover        & 38 & 0.711 & \textbf{0.947} & \textbf{0.947} & 0.737 & 0.605 & 0.632 \\
    paraphrase / alias         & 33 & 0.848 & \textbf{0.970} & \textbf{0.970} & 0.909 & 0.818 & 0.758 \\
    unsupported intent         & 25 & \textbf{0.920} & 0.720 & 0.640 & 0.720 & 0.760 & 0.640 \\
    \bottomrule
  \end{tabular}
\end{table}

The remaining models lie below this frontier. DeepSeek combines
competitive argument accuracy on supported tools with a non-trivial
rate of malformed call strings and hallucinated tools. GPT remains
useful as a conservative reference, ranking second on conflicting
context and competitively on unsupported intent but trailing on
stateful supported requests. Kimi has the lowest accuracy on every
metric except \emph{no tool needed}, where its willingness to
refuse calls reaches 1.000.

\subsection{Free-Form Response Quality}

The benchmark scores tool-calling behaviour but not free-form
text. Under the Zero-Leakage contract, the free-form field is
shown to the user verbatim alongside the structured tool outputs,
and its quality matters most precisely on cases where the model
produces no tool call: there is no template-rendered output to
fall back on, so the user receives only what the model writes.
The 97 no-call cases in the reviewed dataset were therefore the
subject of a complementary evaluation of free-form response
quality.

Each free-form response was scored on a three-point rubric: a
score of two for a response that correctly identifies the
situation, asks for the specific information that is missing, or
explains the limitation that prevents the request from being
satisfied; a score of one for a response that points in the right
direction but is incomplete, for instance offering the catalogue
of available functions in place of a specific clarifying question;
and a score of zero for an off-topic, fabricated, or clinically
inappropriate response, or for a tool call when none was expected.
Scores were normalised to the unit interval. Scoring was carried
out by Claude Opus 4.7 acting as LLM judge over the full set of 576 valid responses (six
models against 97 cases, with three of Kimi's responses excluded
because they failed the response schema and could not be
evaluated). Aggregate results are reported in
Table~\ref{tab:freeform-quality}.

\begin{table}
  \caption{Free-form response quality on the 97 no-call cases,
    scored 0--2 by an LLM judge and normalised to $[0,1]$. CC:
    conflicting context. MRA: missing required argument. NTN: no
    tool needed. UI: unsupported intent.}
  \label{tab:freeform-quality}
  \small
  \begin{tabular}{lccccc}
    \toprule
    Model & CC & MRA & NTN & UI & Overall \\
    \midrule
    Gemma~4 MoE   & \textbf{0.87} & 0.93          & \textbf{0.98} & \textbf{0.96} & \textbf{0.94} \\
    Gemma~4 dense & 0.20          & \textbf{0.94} & 0.96          & 0.84          & 0.80 \\
    Gemini~3.1~flash    & 0.13          & 0.90          & 0.91          & 0.80          & 0.76 \\
    DeepSeek V4   & 0.53          & 0.74          & 0.91          & 0.76          & 0.75 \\
    Kimi K2.5     & 0.40          & 0.75          & 0.93          & 0.79          & 0.75 \\
    GPT~5.4~mini  & 0.57          & 0.69          & 0.83          & 0.80          & 0.73 \\
    \bottomrule
  \end{tabular}
\end{table}

The free-form ranking partially mirrors the tool-calling ranking
but is not redundant with it. Gemma~MoE leads on both metrics.
Gemma~dense ranks second on free-form quality despite having the
lowest no-call accuracy of the leading models: when it does
abstain its explanations are competitive, but it abstains less
often than Gemma~MoE in conflicting-context cases. Gemini
exhibits the inverse pattern: a respectable no-call accuracy
paired with the lowest score on conflicting-context text, because
the model overcalls on these cases almost universally and the few
abstentions it produces are not of high quality. Three failure
modes recur across the no-call cases. The first is arbitrary
argument substitution, in which a model defaults to
\texttt{patient\_id=0} or \texttt{patient\_id=1} without
disclosure when the patient identifier is not specified; GPT and
DeepSeek exhibit this behaviour, the two Gemma variants do not.
The second is deflection to the catalogue of available functions
in place of a clarifying question. The third is helpful overreach
on out-of-scope clinical requests, in which a model treats a
question about treatment, prognosis, or referral as a
data-retrieval task and chains together patient-level explanation
calls in response. Gemma~MoE consistently declines such requests
and offers to redirect to the supported functions; other models
decline inconsistently, and Kimi was observed to fabricate a
non-existent tool name in one case.

\subsection{Latency and Selection}

Latency was measured on the same evaluation runs. Gemma~MoE has a
median response time of 2.0\,s with a 90th-percentile of 4.0\,s
and a maximum of 32.7\,s. DeepSeek V4 Flash has a median of
3.5\,s, a 90th-percentile of 12.2\,s, and a maximum of 134.2\,s.
The Gemma~MoE latency profile is the more suitable of the two for
an interactive deployment.

Taken together, the tool-calling and free-form evaluations
identify Gemma~4 Mixture of Experts as the leading default for
the heart deployment under the Zero-Leakage contract: the model
leads on overall joint goal accuracy, holds a substantial
advantage on no-call discipline and on free-form response quality,
and has the most favourable latency profile among the candidates
with latency reported. Gemma~4 dense and Gemini~3.1 Flash Lite
are competitive alternatives in deployments that prioritise
stateful supported-tool execution over conservative ambiguity
handling. The user study reported in Section~\ref{sec:user-study} was conducted 
with Gemini~3.1 Flash Lite rather than Gemma~4 MoE. The choice was driven by 
deployment constraints during the study period: Gemini provided greater API 
stability, lower operational latency under sustained interactive use, 
and simpler integration within the existing production environment 
used for participant access.

\section{User Study Design}
\label{sec:user-study}

The user study is designed as a within-subjects comparison between
the Explainability Assistant and the dashboard, with each
participant completing parallel task blocks on both interfaces.

\subsection{Participants}

Of the ten participants recruited, five completed the study 
appropriately and contribute to the analyses reported below; 
the remaining five submissions were excluded during quality 
control (see Section~\ref{sec:limitations} for the exclusion 
criteria and their justification).
The pool spans healthcare professionals
and trainees with mixed clinical and public health backgrounds,
including a radiotherapist, a final-year medical student, public
health and physiotherapy students at master's level, and an MPH
student with an epidemiology focus. Self-reported familiarity with
machine learning ranges from minimal exposure to intermediate
working knowledge; no participant identified as a machine learning
specialist.



\subsection{Compared Interfaces}

Two interfaces are compared, both providing access to the same
Decision Tree classifier and the same set of post-hoc explanation
methods (SHAP, DiCE, what-if analysis) described in
Section~\ref{sec:task}. The first is the Explainability Assistant,
which exposes the explanation methods through natural-language
dialogue. The second is the Explainer
Dashboard \cite{oege_dijk_2023_7633294},
an open-source dashboard widely used as a reference point for
interactive explanation interfaces, providing tabbed access to
SHAP feature importance plots, individual prediction inspection,
global feature importance rankings, and interactive sliders for
what-if analysis.

\subsection{Tasks and Subjective Evaluation}

Each interface is paired with a parallel task block of six
questions: retrieving information about the dataset or model artefact,
retrieving an individual patient's prediction, identifying the
feature that contributed most to the prediction,
evaluating the feature's meaningfulness to the user, evaluating a
hypothetical change to one feature, finding additional scenario
of model prediction change.
The blocks are matched in structure and difficulty but
use different patient records to prevent memorised answers from one block
carrying over to the other. Participants could not leave
a question blank and were instructed to write explicit ``I cannot tell'' 
if the interface does not allow them to find an answer within
reasonable effort, so that the task accuracy measure reflects
what the interface enabled the participant to discover rather
than what the participant could guess.

After each task block, participants complete a five-item subjective
evaluation on a five-point Likert scale (1 = strongly disagree,
5 = strongly agree) covering ease of use, trust in the outputs,
clarity of the explanations provided, self-reported understanding
of how the system operates, and likelihood of using the interface
in future work. A final open-ended item invites free-text feedback
on the interfaces used.

\subsection{Procedure}

The study was administered online and asynchronously through
Google Forms, allowing participants to complete the survey at
their own convenience and without researcher supervision. Each
participant received a unique access code linking their responses
to the appropriate conversation for analysis. The questionnaire was
organised into six sequential sections: background information,
Task Block A, post-task evaluation A, Task Block B, post-task
evaluation B, and open-ended final feedback. Interface order was
counterbalanced across participants, partially mitigating the
ordering effect in which familiarity acquired during the first
block can inflate performance and usability ratings for the
second.

Participants received a 25\,EUR gift card for the questionnaire
on the condition that the task was completed properly, and the 
criteria used to judge proper completion (and the exclusions 
that resulted) are described in Section~\ref{sec:limitations}.
Participants who indicated willingness to take
part in a follow-up interview received a separate 25\,EUR payment
by bank transfer after the interview was completed. The two
compensation streams are kept distinct so that the questionnaire
payment is not contingent on subsequent interview availability,
and the interview payment is not used as leverage on
questionnaire content.

\subsection{Follow-up Interviews}

Three of the five valid participants additionally took part in
follow-up interviews of approximately twenty minutes each,
conducted by one of the authors over Zoom. An interview script
guides the conversation, but in practice the sessions take the
form of open discussions, with the script serving as a checklist
of topics rather than a rigid sequence. Topics addressed included
the participant's reasoning strategies during the task blocks,
points of friction with each interface, perceived limits of
either system, and which interface the participant would prefer
to use in clinical or research practice and why. With the
participant's explicit consent, each session was recorded in
audio and video, and the Zoom AI Companion note-taking feature
was used to produce a written summary. The recordings served as
the primary record from which transcripts could be produced; 
the automatically generated summaries supported quicker
inspection during analysis. The interview component was designed
primarily to inform the future development of both interfaces,
with conversations focused on what either system would need to
offer to be usable in clinical practice and on the design
changes participants believe would close the gap between a
research prototype and a tool fit for routine use.

\section{Results}
\label{sec:results}

The results are reported in three parts.
Section~\ref{sec:quantitative} summarises the quantitative
findings from the questionnaire.
Section~\ref{sec:freetext-feedback} reports the analysis of the
free-text responses provided alongside the task questions and in
the final open-ended item.
Section~\ref{sec:interview-findings} reports the findings from
the follow-up interviews. Across all three layers a single
convergent finding emerges: participants did not regard the two
interfaces as substitutes for one another and recommended their
integration into a single clinical tool.

\subsection{Quantitative Findings}
\label{sec:quantitative}

This subsection reports objective task accuracy and the Likert
ratings collected after each task block, for the participants whose
responses were retained after the validity checks described in
Section~\ref{sec:user-study}. Given the size of the sample, the
analysis is confined to descriptive comparison and median paired
differences, with inferential claims avoided in favour of
effect-pattern observations.

\paragraph{Task accuracy.}
On the questions with an unambiguous ground-truth answer, the
conversational assistant produced more uniformly correct responses
than the dashboard. The patient-level prediction question
and the hypothetical single-feature-change question were
answered correctly by every participant on both interfaces,
indicating that point predictions and simple counterfactual
sensitivities were communicated reliably by either modality.
Differences between interfaces appeared on the model-description,
feature-importance, and exploration questions.

On the chat-side model-description question, four of the five
participants returned a correct description of the underlying
Decision Tree classifier; the fifth identified the language model
providing the conversational interface itself in place of the
classifier, a failure mode discussed further in
Section~\ref{sec:freetext-feedback}. On the dashboard the parallel
dataset-description question elicited two ``can't find'' responses
and one fragmentary answer, with the remaining two participants
returning a substantive description.

The clearest accuracy contrast appeared on the feature-importance
question. For the chat-side query identifying the strongest
contributor to the prediction for patient~ID~39, four of the five
participants returned the local SHAP attribution (\emph{trestbps})
that the conversational assistant emits directly. For the
dashboard-side query on patient~55 all five participants instead
returned the top global feature (\emph{thal}), with two of them
flagging the local-versus-global ambiguity explicitly. The contrast
is consistent with the conversational interface scoping its
response to the entity under discussion and the dashboard
juxtaposing global and local views in adjacent panels without
disambiguating them (Section~\ref{sec:freetext-feedback}).

The dashboard scenario question, which asks participants to
identify a feature modification that flips the prediction, drew
one explicit abandonment of the partial dependence plot and four
substantive responses, ranging from a single feature change to a
sequence of modifications that drove the predicted probability of
disease to its maximum. The corresponding chat-side modification
question produced substantive responses in every case, including
responses that used the counterfactual function to generate
ranked suggestions automatically.

\begin{figure*}[h]
    \centering
    \includegraphics[width=1\linewidth]{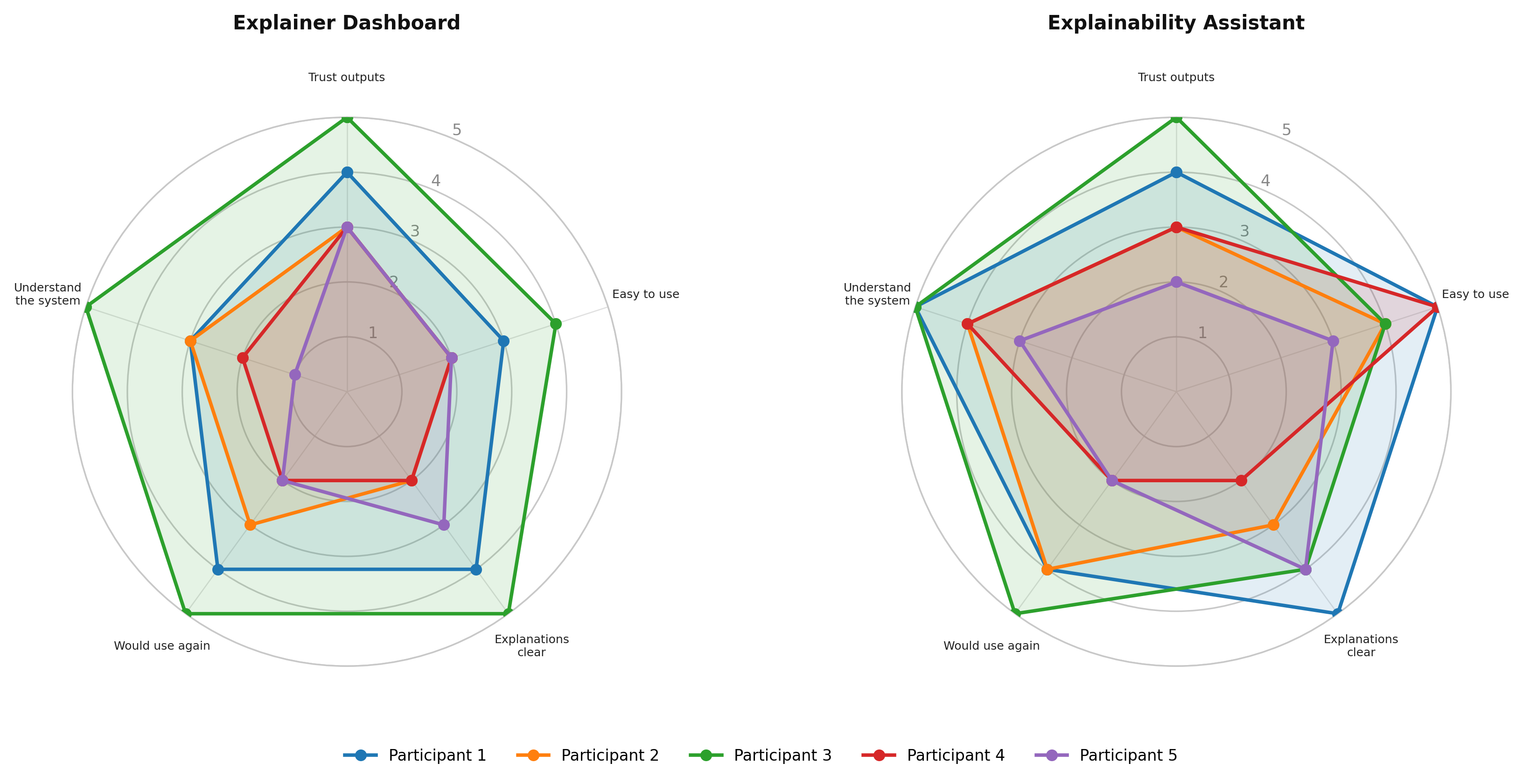}
    \caption{Subjective evaluation of the two interfaces ($n=5$;
    $1 = $ strongly disagree, $5 = $ strongly agree). Individual
    participant ratings comparing the Explainer Dashboard (left)
    and the Explainability Assistant (right) across five
    dimensions.}
    \label{fig:radar}
\end{figure*}

\paragraph{Likert ratings.}
Each interface received a five-item subjective evaluation on a
five-point scale covering ease of use, trust in the outputs,
clarity of the explanations provided, self-reported understanding
of how the system operates, and likelihood of future use.
Individual participant ratings are reported in
Figure~\ref{fig:radar}.

The conversational assistant carries a substantial advantage on
ease of use, clarity, and self-reported understanding, with a
median paired difference of two Likert points on each of these
dimensions. Trust in outputs and likelihood of future use are at
parity at the median. Within-participant comparisons preserve the
same direction: every participant rated the conversational
assistant at least as easy to use as the dashboard and at least
as understandable, with the largest within-participant difference
on ease of use reaching three Likert points. The single
participant with the most substantive clinical experience rated
the two interfaces equivalently on every dimension shown in
Figure~\ref{fig:radar}, suggesting that the consistent ease-of-use
and understanding advantages observed across the rest of the pool
are concentrated among participants with less prior exposure to
dashboard-style explanation tools.

The sample is too small to support inferential claims with
confidence, and the descriptive pattern reported here should be
read as an effect-direction signal rather than as a measurement
of population-level preferences. The pattern is consistent with
the interpretation that the conversational modality reduces
interaction cost for the kinds of clinical inquiry the task block
exercises, without yet establishing a corresponding trust or
adoption advantage. The reasons participants gave for their
reactions are taken up in Section~\ref{sec:freetext-feedback} and
Section~\ref{sec:interview-findings}.

\subsection{Free-Text Feedback and Task Engagement}
\label{sec:freetext-feedback}

The questionnaire generated qualitative data on two further
layers: the textual answers participants gave to the task
questions themselves, several of which include written
explanations of how the answer was reached, and the final
open-ended item asking for general feedback. Five recurring
patterns emerged across these responses.

\paragraph{Honest abandonment in place of guessing.} Several
participants opted out of questions when the interface did not
let them locate the answer within reasonable effort, rather than
producing a confident-sounding response without grounds. Two
participants returned ``I can't find'' in place of a description
of the dataset on the dashboard, and one wrote that they were
going to ``give up'' on the dashboard's partial dependence plot,
which they had found ``too inconvenient'' for the question. The
instruction that blanks were preferred over guesses functioned
as intended: the resulting blank rate is itself a measure of how
often each interface failed to support a question rather than of
participant capability.

\paragraph{Confusion of the conversational layer with the
classifier.} A failure mode specific to the Explainability
Assistant emerged in answers to the model-description question.
One participant identified the underlying predictive model as
``gemini 3.1 flash lite'', which is the language model providing
the conversational interface rather than the Decision Tree
classifier on which all predictions are computed. The conflation
is consistent with the surface presentation of the assistant: the
LLM is the most visible component, the classifier is invoked by
tools the user does not directly observe, and the distinction is
never explicitly surfaced in the chat. The implication for
clinical deployment is that an interface relying on conversational
mediation should be transparent about which component performs
the inference and which mediates the conversation, since a user
who attributes the prediction to the LLM is in a poor position
to reason about the model's reliability.

\paragraph{Local versus global feature importance.} The
dashboard's response to the question of which feature contributed
most to a specific patient's prediction was characterised by a
recurring uncertainty about whether the answer should be drawn
from the local explanation for that patient or from the global
feature ranking for the model as a whole. Two participants
recorded the ambiguity directly. One wrote ``Thalassemia? Or
average of population? That's unclear for me'', and another
flagged ``thal = 3.0 (average of population should probably be
excluded as the starting point)''. The convergence of independent
responses on the same uncertainty suggests that the dashboard's
juxtaposition of global and local feature plots does not reliably
communicate which view applies to a question phrased at the level
of an individual patient. Responses through the conversational
assistant to the parallel question on a different patient did
not exhibit the same pattern, with most participants returning
the local SHAP attribution directly. The contrast is consistent
with the conversational modality scoping its response to the
entity under discussion, while the dashboard presents both views
in adjacent panels and leaves disambiguation to the user.

\paragraph{Local clinical context and unit mismatch.} The final
open-ended item drew substantive responses on the integration of
the tools into specific clinical workflows. One participant
working in primary care in Estonia observed that general
practitioners in that setting do not have access to the \emph{ca}
feature, which limits the practical relevance of the model in
that context, and added that Estonian clinical practice records
cholesterol in mmol/L rather than the mg/dl displayed by the
interface, so that the figures shown required mental conversion
before they could be interpreted at a clinical glance. A second
participant phrased a related concern in writing that a
``medical practitioner would want to make min amount of klicks''.
These responses point to a class of design issues that originate
not in the explanation methods but in the assumed clinical
context: the input feature set, the display units, and the
interaction cost are calibrated to the dataset's source
population and to a particular professional category, and
neither interface currently exposes configuration affordances
for adapting them.

\paragraph{Volunteered design suggestions.} The free-text
responses included concrete design proposals from three
participants, clustering around the dashboard's presentation 
rather than its explanation content: plain-language tooltips for 
abbreviated feature labels, clinical units and reference ranges 
next to feature inputs, a reset-patient control, ranges rather 
than single values for categorically presented features, and 
clearer feedback when a what-if modification changes the 
underlying prediction. The convergence of these suggestions on
tighter coupling between the user's modification action and the
visible model output indicates that the dashboard's strongest
perceived weakness in this sample is not the explanation methods
themselves but the rendering of state changes in the explanation
interface.

\subsection{Interview Findings}
\label{sec:interview-findings}

The follow-up interviews extend several patterns observed in the
questionnaire responses and reinforce the convergent finding
that participants do not regard the two interfaces as substitutes.
Findings are derived from analysis of the audio and video recordings,
supported by the AI-generated Zoom meeting summaries as a cross-reference;
the recordings served as the authoritative record where summaries 
were ambiguous or incomplete. The three interview participants are
referred to as P1 (physiotherapy MSc graduate), P2 (public health
MSc student), and P3 (radiotherapy MSc student).

\paragraph{Integration over selection.} All three participants,
when asked to express a preference, recommended integrating the
conversational and dashboard modalities rather than adopting one
to the exclusion of the other. P1 proposed embedding the
conversational assistant as a chatbot panel within the dashboard
interface; P2 framed the integration in the opposite direction,
suggesting that dashboard elements be embedded into the
conversational assistant; P3 endorsed the integration position
more agnostically, treating both modalities as legitimate and
complementary. The convergent point across the three formulations
is the same: a clinical XAI tool should support both interaction
modalities together. P1 additionally observed that the
conversational assistant tended to return ranked tables rather
than direct answers when asked for the single most important
factor in a prediction, a calibration choice consistent with the
participants' preference for visual dashboard elements that
present information in structured rather than enumerated form.

This finding contrasts with patterns reported in the prior
conversational XAI literature, in which evaluations of
natural-language interfaces have generally found that
conversational modalities are preferred over dashboards by users 
with no expertise in machine learning \cite{slack2023,samimi2025}. 
In the clinical
setting examined here, the conversational modality's ergonomic
advantages are recognised but do not translate into a
recommendation to displace dashboards.

\paragraph{Conversation continuity and question structuring.} P1
and P2 independently identified the absence of conversation
history and question categorisation as the sharpest usability
gaps in the conversational assistant. Both explicitly drew
analogies to consumer LLM interfaces (ChatGPT, Claude) when
describing what they wanted: a visible record of prior questions,
the ability to copy and modify previous prompts rather than
retype them, and a structuring of questions that distinguishes
patient-specific queries from broader inquiries about the model
or dataset. P1 added that question categorisation should be
performed automatically rather than by requiring users to select
a category before asking, since explicit categorisation steps
were perceived as friction. The convergence on these specific
UI features suggests that contemporary user expectations of
conversational interfaces are shaped strongly by the consumer
LLM market, and that clinical tools failing to meet these
expectations will read as unfinished regardless of their
analytical capabilities.

\paragraph{Onboarding friction is uniform but locally specific.}
All three participants reported difficulty during initial use,
but the location of the difficulty differed and tracked
participant background. P1, whose training is in physiotherapy
and includes no formal statistics curriculum, identified
statistical terminology as the central barrier and recommended
interactive tutorials and clearer plain-language explanations of
metrics. P2, with a public health background including
statistical training, did not raise this concern but identified
instead the chat interface's lack of conversation overview and a
confusing dropdown control on the dashboard. P3 reported
difficulty with graph generation in the dashboard, particularly
pie chart creation, while finding the prediction content easy to
understand. The implication is that onboarding effort cannot be
reduced to a single intervention: clinical users with different
training backgrounds encounter friction in different layers of
the same interface.

\paragraph{The dashboard signals professional expertise.} A
distinctive observation from P3 is that a dashboard interface
with graphical elements is more appropriate for professional
clinical settings than a pure conversational interface, on the
grounds that the dashboard demonstrates the operator's knowledge
and expertise. This observation is sociological rather than
ergonomic: it concerns what the tool communicates about the
person using it, not whether the tool is easy to use.
Conversational interfaces, however accessible, may carry
connotations of low-stakes consumer software that conflict with
professional clinical identity, and integrated tools may need to
preserve dashboard-style visualisation not only for information
communication but for the professional framing of the clinical
interaction itself. This observation was raised by only one
participant and is reported as a hypothesis worth testing in
larger samples rather than as a generalised finding.

\section{Discussion}
\label{sec:discussion}

The most consistent finding across the quantitative measures, the
free-text questionnaire responses, and the follow-up interviews is
that the comparison did not yield a clear preference for either
modality. The study should be interpreted as an exploratory formative 
evaluation rather than as a statistically powered comparative trial. 
With five valid participants and three follow-up interviews, the aim was 
not population-level generalisation or significance testing, but the 
identification of robust design patterns, recurring usability frictions, 
and convergent preferences across quantitative and qualitative evidence. 
For this reason, descriptive comparisons and qualitative convergence are 
treated as the primary source of evidence, while Likert differences and 
task accuracy are interpreted as effect-direction signals rather than inferential results. 
The value of the study lies in clarifying what participants consider necessary 
for clinical usability and in informing the design of the next prototype iteration. 

The Likert ratings and task accuracy figures reported in
Section~\ref{sec:quantitative} do not separate the two interfaces
along a single dimension, and the qualitative analyses
(Sections~\ref{sec:freetext-feedback} and \ref{sec:interview-findings})
converge on a more substantive observation: participants treated
the conversational and dashboard modalities as complementary rather
than competing, and all three interview participants explicitly
recommended an integrated tool combining the two. This pattern
contrasts with earlier conversational XAI evaluations, where users 
with less exposure to machine learning have generally preferred natural-language
interfaces over dashboards \cite{slack2023,samimi2025}, and is
consistent with the more recent observation that conversational
modalities are recognised for their ergonomic advantages but do
not, on their own, displace the dashboard view in settings where
both modalities are available together.

The qualitative analyses are informative about what an integrated
tool would have to preserve from each modality. From the
conversational side, participants valued the natural-language entry
point and the fact that the assistant scoped its response to the
entity under discussion, which avoided the local-versus-global
ambiguity that recurred in the dashboard responses to questions
about individual patient predictions. From the dashboard side,
participants valued the simultaneous availability of global and
local feature views, the visual reference values, and (in one
participant's framing) the professional appearance of a graphical
interface in a clinical setting. The two sets of strengths are
complementary rather than redundant: conversational scoping
addresses a specific weakness of the dashboard, while the
dashboard's visual reference frame addresses a specific weakness
of the conversational assistant when participants asked questions
that the LLM answered with ranked tables rather than with direct
answers.

The free-text responses surfaced a related set of design
implications that an integrated tool would need to address. The
conflation of the conversational layer with the underlying
classifier, observed in at least one participant's answer to the
model-description question, indicates that an interface relying on
LLM mediation should be transparent about which component performs
the inference and which mediates the conversation. The clinical
context observations, including the absence of the \emph{ca}
feature in Estonian primary care and the cholesterol unit mismatch
between mg/dl and mmol/L, point to a class of issues that originate
not in the explanation methods but in the assumed clinical context,
and that would need to be addressed through configuration
affordances that neither interface currently exposes. The
dashboard-specific suggestions volunteered by participants,
including plain-language tooltips for abbreviated feature labels,
clinical units and reference ranges next to feature inputs, a
reset-patient control, and a clearer coupling between what-if
modifications and the visible prediction state, all point to the
same underlying issue: the dashboard's strongest perceived weakness
in this sample is not the explanation methods themselves but the
rendering of state changes in the explanation interface.

The trust calibration concerns reviewed in
Section~\ref{sec:related-work} bear on these findings in a specific
way. The empirical literature has documented that explanation
provision can promote over-reliance through false-confirmation
effects \cite{informatics12040135}, and the conversational modality has
been identified as a particular risk because the persuasive
fluency of natural-language responses can mask the limits of the
underlying model \cite{he2025}. The conflation of the LLM with the
classifier in the free-text responses is consistent with this
risk and suggests that the visual separation
of model and interface in the dashboard contributes something to
trust calibration that the conversational modality does not
currently match. A natural design response is an integrated tool that combines a
conversational entry point with a clear visual separation between
the LLM and the underlying classifier, so that users trust (or
distrust) the right component. Whether such an integration would
in fact produce better alignment between user trust and model
reliability is an open empirical question, and not one that can
be settled by the fact that each modality has some desirable
property on its own.

\section{Conclusion}
\label{sec:conclusion}

This paper reported a within-subjects user study comparing the
Explainability Assistant, a conversational XAI system based on LLM
function-calling, with a traditional dashboard interface for the
explanation of a heart disease classification model. Both
interfaces were configured to expose the same Decision Tree
classifier and the same set of post-hoc explanation methods
(SHAP feature attribution, DiCE counterfactuals, and what-if
analysis), so that the comparison isolated the contribution of
the interaction modality rather than confounding it with
differences in explanation content.

Five participants with mixed healthcare backgrounds completed the
study, and three additionally took part in follow-up interviews.
The quantitative measures (task accuracy and Likert ratings) did
not produce a clear preference for either modality. The
qualitative analyses, drawing on free-text questionnaire responses
and the interview transcripts, converged on a more substantive
finding: participants treated the conversational and dashboard
modalities as complementary rather than competing, and all three
interview participants explicitly recommended an integrated tool
combining the two. This pattern is at odds with earlier
conversational XAI evaluations, where non-expert users have
generally preferred natural-language interfaces over dashboards.
In the clinical setting examined here, the conversational
modality's accessibility advantages were recognised but did not
translate into a recommendation to displace the dashboard.

The natural direction for future work is a prototype of the integrated tool
implied by these findings: a single clinical interface that
combines a conversational entry point with the dashboard's visual
representation of the underlying model and its explanations.
Such a prototype would convert the qualitative recommendation
into a concrete artefact that can itself be evaluated, and would
make it possible to test empirically whether the integration of
two modalities inherits the strengths of each or introduces new
trade-offs of its own.
Findings suggest that conversational XAI should not be evaluated 
as a replacement for dashboards, but as a complementary modality 
whose value emerges through integration.

\section{Limitations}
\label{sec:limitations}

Four limitations constrain the interpretation of the present study. 
First, the participant pool is small: of ten recruited participants, 
five responses were excluded during quality control because the 
questionnaires showed indications of malicious AI assistance or 
other unethical participation motivated by the financial compensation, 
leaving five valid questionnaire responses and three follow-up 
interviews. Inferential statistical analysis is therefore not 
attempted, and the quantitative findings are reported as descriptive 
effect patterns. Second, the study was conducted asynchronously and 
without supervision, which improved accessibility but required 
retrospective detection of low-quality responses; supervised or 
observational protocols would raise the effective response rate in 
future replications. Third, the evaluation covers a single clinical 
task, dataset, and interpretable classifier, and the findings may 
not transfer directly to higher-dimensional data or less inspectable 
models. Fourth, the dashboard baseline is one implementation among 
several, and the conversational interface was evaluated with a 
single deployed LLM configuration; alternative choices on either 
side may shift the specific pattern of strengths and weaknesses, 
though the integration finding is at the level of modality rather 
than of implementation.

\section{Resources}
\label{sec:resources}

To support reproducibility, the Explainability Assistant
deployment used in the study, together with the questionnaire
and follow-up interview protocol, have been prepared as
supplementary materials.

These materials include:
(i) a live demonstration of the heart disease deployment used
by participants in the user study, exposing the same Decision
Tree classifier and explanation functions described in this
paper; and
(ii) the full questionnaire and interview script used for
data collection.

\begin{enumerate}
    \item Explainability Assistant website: \\ \href{https://explainabilityassistant.com}{https://explainabilityassistant.com}

    \item Explainability Assistant, back-end application code: \\ \href{https://github.com/krkv/explainability-backend}{https://github.com/krkv/explainability-backend}

    \item Explainability Assistant, front-end application code: \\ \href{https://github.com/krkv/explainability-frontend}{https://github.com/krkv/explainability-frontend}

    \item Gold parse datasets, dataset generation prompts and scripts, LLM evaluation scripts: \\ \href{https://github.com/krkv/llm-evaluation}{https://github.com/krkv/llm-evaluation}

    \item User survey and interview questions: \\ \href{https://docs.google.com/document/d/1-OuLOxyPnxfWKDosSZ5IiwM-qY_jbHyKKynxvMC84oo/edit?usp=sharing}{https://tinyurl.com/4m2b4m2n}
\end{enumerate}

\section*{Acknowledgments}
This work was supported by the Estonian Research Council grant PRG2006.
The authors thank the participants of the user study.

\section*{Declaration on Generative AI}

During the preparation of this work, the authors used generative AI tools, including ChatGPT and Claude, to assist with language refinement, clarity improvements, structural suggestions, grammar and spell checking, and the identification of sections requiring further revision or strengthening. These tools were used to support drafting and editing, but not for the generation of the scientific contributions, experimental results, data analysis, or final research conclusions.

All conceptual decisions, methodological choices, experimental design, interpretation of results, and final manuscript content remain the full responsibility of the authors. All generated suggestions were critically reviewed, revised where necessary, and verified by the authors before inclusion in the final manuscript.

\bibliographystyle{IEEEtran}
\bibliography{explimed_paper}

\end{document}